\documentclass[superscriptaddress,amssymb,twocolumn,pre]{revtex4-2}

\usepackage{natbib,graphicx,dcolumn,bm,amsmath,color,bm,subfigure}

\newcommand{\eq}[1]{Eq.~(\ref{#1})}

\newcommand{\fig}[1]{Fig.~\ref{#1}}

\definecolor{amber}{rgb}{1.0, 0.75, 0.0}

\newcommand{\red}[1]{ {\color{red} #1}}

\newcommand{\eeq}{ \end{equation} }
\newcommand{\beq}{ \begin{equation} }

\newcommand{\eea}{ \end{align} }
\newcommand{\bea}{ \begin{align} }

\newcommand{\bhu}{{\bf \hat{u}}}

\newcommand{\bx}{{\bf \hat{e}_{1}}}
\newcommand{\by}{{\bf \hat{e}_{2}}}
\newcommand{\bz}{{\bf \hat{e}_{1}}}

\newcommand{\bhe}{{\bf \hat{e}}}

\newcommand{\br}{ {\bf r }}

\newcommand{\bn}{ {\bf \hat{n} }}

\begin{document}

%\title{Screwed up with a helix} 

\title{Spontaneous chiral symmetry breaking in polydisperse achiral near-rigid nematogens}

% Force line breaks with \\

\author{W. S. Fall}
\email{william.fall@cnrs.fr}
\affiliation{Laboratoire de Physique des Solides - UMR 8502, CNRS,  Universit\'{e} Paris-Saclay, 91405 Orsay, France}

\author{H. H. Wensink}
\email{rik.wensink@cnrs.fr}
\affiliation{Laboratoire de Physique des Solides - UMR 8502, CNRS,  Universit\'{e} Paris-Saclay, 91405 Orsay, France}

\begin{abstract}
Understanding chirality transfer from the molecular to the macroscopic scale poses a significant challenge in soft and biological condensed matter physics. Many nanorods of biological origin not only have chiral molecular features but also exhibit a spread in contour length leading to considerable size dispersity.    On top of this, random backbone fluctuations are ubiquitous for non-rigid particles but their role in chirality transfer remains difficult to disentangle from that of their native chirality imparted by their effective shape or surface architecture.  We report spontaneous entropy-driven chiral symmetry breaking from molecular simulations of cholesteric liquid-crystals formed from achiral   bead-spring rods with a continuous spread in contour length and marginal chain bending. The symmetry-breaking is caused by long-lived chiral conformations  of long rods undergoing  chiral synchronization leading to a homochiral twisted nematic. A  simple  theory demonstrates that even without chiral synchronization, the presence of  shape-persistent configurational fluctuations along with length-dispersity can be harnessed to generate non-zero chirality at moderate polydispersity.
\end{abstract}

\date{\today}

\maketitle

The discovery of chiral liquid crystal (LCs) phases over 150 years ago has had
profound scientific and industrial implications \cite{wade2020natural,planer1861notiz,reinitzer1888beitrage}. By virtue of their long-ranged orientational order and helical mesostructure these phases have found widespread use  in optoelectronics and display technology \cite{bisoyi2014light,pieraccini2011chirality} and continue to inspire fundamental researchers for the fascinating structures they may form \cite{wu2022hopfions, gibaud2012reconfigurable}. 
A broad distinction can be made between thermotropic LCs where chirality is  encoded by chemical groups transmitting chiral intermolecular torques and lyotropic LCs of rod-shaped colloids, often of biological origin, which  are known to form chiral superstructures induced by their effective chiral shape (e.g. a helix) or surface pattern. Prominent examples of the latter class are nanocrystalline chitin \cite{revol1993vitro} or cellulose \cite{lagerwall2014cellulose, tran2020understanding},  amyloid fibrils \cite{nystrom2018liquid,jin2024structural}, filamentous virus  \cite{grelet2003origin}, or mineral moieties such as carbon nanotubes \cite{artyukhov2014nanotubes}.     
 In most cases, the chirality of the constituents is ascribed to some net chiral  shape and unravelling the bottom-up transfer of chirality across length scales has been the subject of considerable  investigation  for  biopolymers  LCs \cite{cecconello2019controlling, parton2022chiral, fittolani2022bottom, grelet2024elucidating},  topological polymers \cite{lukin2005knotting, zhao2023can}, living matter \cite{mitov2017cholesteric} and indeed for nanotechnology at large \cite{morrow2017transmission}.  One surprising outcome is that spontaneous chirality may also  emerge from molecular shapes that are strictly achiral but have some broken particle symmetry, such as banana-shaped molecules \cite{dozov2001spontaneous, hough2009chiral} or from molecules with transient chiral features \cite{tschierske2016mirror}.  For many biopolymers, however, the notion of a uniquely identifiable (effective) particle shape, be it chiral or non-chiral, is severely complicated by two factors. The first is the presence of backbone fluctuations, which are to be accounted for when the rods are not entirely rigid but undergo small conformational transformations.    Most bio-colloidal rods fall into the regime of so-called marginal flexibility where the persistence length $L_{P}$, a measure for the length scale over which local backbone fluctuations are correlated, exceeds the contour length $L$ of the rods but remains finite nevertheless ($L_{P}$  is infinity  for rigid rods). Recent studies have established the role of backbone fluctuations  in being rather crucial in reinforcing chiral transmission between non-rigid biological mesogens that are natively chiral such as DNA origami \cite{tortora2020chiral} and  filamentous virus rods \cite{grelet2024elucidating}.  The second complicating factor is that most biopolymers exhibit a considerable spread in  length and diameter. Size dispersity is a key but often overlooked feature of widely studied systems such as cellulose nanocrystals \cite{jakubek2018characterization} and the  relation between size dispersity and chirality transfer turns out rather subtle even for rigid rods \cite{wensink2019effect,sewring2023effect} while the additional impact of backbone fluctuations remains elusive to date.  
 
\begin{figure}[htb]
\includegraphics[width=\columnwidth]{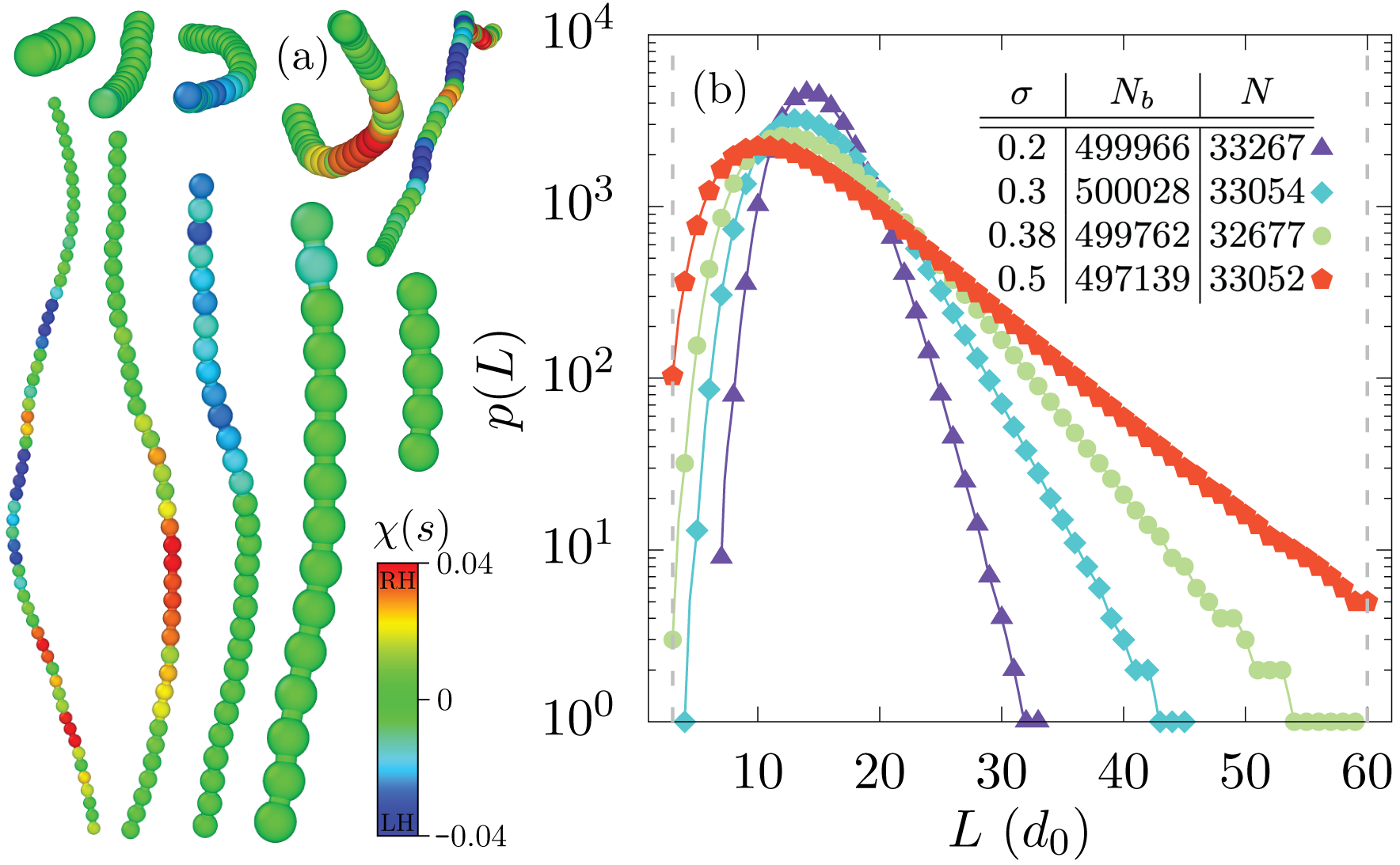}%
\caption{\label{fig:dist} (a) Snapshots of near-rigid bead-spring rods of different lengths but identical diameter comprising 5, 15, 30, 40 and 59 monomers, respectively. The rods are color-coded according to their local chirality $\chi$ [\eq{intrachiral}] as a function of increasing contour length $s$ taken w.r.t the first bond vector. (b) Discretized log-normal rod length distributions $p(L)$, with average rod aspect ratio $ \bar{L} /d_{0}=15$, where dashed lines indicate short and long-rod cut-offs.  }
\end{figure}

 In this letter, the combined impact of shape fluctuations and polydispersity is addressed using a model which is fundamentally non-chiral in its ground state, namely cylindrical rods with marginal backbone flexibility (``floppy" cylinders,  \fig{fig:dist}a).  To that end, a minimal model is employed where rods are represented as bead-spring chains with spherical beads of diameter $d_{0}$. The rods are held together by harmonic bonds of the form $U_{\mathrm{bond}}(l)=\frac{1}{2}k_{\mathrm{bond}}(l-l_{0})^{2}$
where $U_{\mathrm{bond}}(l)$ is the potential energy change associated with deforming the bond with length $l$ from its equilibrium separation $l_{0}=d_{0}$ and $k_{\mathrm{bond}}=1000~k_\mathrm{B}T/d_{0}^{2}$ is the spring constant.   Additional harmonic angular potentials are enforced between every 3 beads along the colloid which are defined as follows $U_{\mathrm{angle}}(\theta)=\frac{1}{2}k_{\mathrm{angle}}(\theta-\theta_{0})^{2}$
where $U_{\mathrm{angle}}(\theta)$ is the potential energy change associated with deforming the angle away from its equilibrium angle, $\theta_{0}=180^\circ$ and $k_{\mathrm{angle}}=120~k_\mathrm{B}T/\mathrm{rad}^{2}$ is the spring constant.

Non-bonded interactions between rods take place via a standard WCA potential 
\begin{eqnarray}
U_{\mathrm{LJ_{12-6}}}=4\epsilon\bigg[\Big(\frac{d_{0}}{r}\Big)^{12}-\Big(\frac{d_{0}}{r}\Big)^{6}\bigg],\ \ \ r\leq r_{\mathrm{c}}
\label{e:nonbond}
\end{eqnarray}
where $\epsilon=k_\mathrm{B}T$ denotes the depth of the potential well in terms of the thermal energy with temperature $T$ and Boltzmann's constant $k_\mathrm{B}$,  $r$ the variable inter-bead separation. The potential is cut and shifted at the minimum $r_{\mathrm{c}}=2^{\frac{1}{6}}d_{0}$, ensuring only steeply repulsive interactions such that all phase transformations are entirely entropy-driven. Most importantly, even though some rod conformations may be transiently chiral, e.g. locally resembling a helix, neither handedness (LH or RH) is favored a priori so that the individual rods are strictly {\em non-chiral}.

Molecular dynamics simulations are performed to study the self-assembly of the bead-spring rods for a series of different length polydispersities defined as  $\sigma= ( \langle  L^{2} \rangle - \langle L \rangle^{2} )^{1/2}/\langle L\rangle = 0.2,0.3,0.38$ and 0.5  which correspond to typical size dispersities of stiff rods  \cite{brinkmann2016correlating,lehman2011evaluating}. Log-normal distributions $p(L)$ are sampled [\fig{fig:dist}b] where dashed lines indicate short and long tail cuts such that the minimum and maximum rod lengths are 3 and 60 beads respectively. This ensures the largest rods are at maximum half the size of the simulation cell. All systems comprised approximately $N_{b}=5 \times 10^{5}$ beads and $N =3.3 \times 10^{4}$ rods (\fig{fig:dist}).
%\begin{table}
%\begin{tabular}{c|c|c}
%$\sigma$ & $N_{b}$ & $N$ \\
%\hline
%\hline
%0.2 & 499966 &  33267 \\
%0.3 & 500028 &  33054 \\
%0.38 & 499762 &  32677 \\
%0.5 & 497139 & 33052 \\
%\end{tabular}
%%\end{table}
Throughout, we define an effective rod concentration $c=\tfrac{\pi}{4} N \bar{L}^{2} d_{0}/V$ with $\bar{L}$ the average rod length and $V$ system volume. 
From the worm-like chain model we estimate the persistence length $L_{P} = B/k_{{\rm B}}T =  240 d_{0}$ with $B = 2 d_{0} k_{\rm angle}$ the bending modulus of the rods.  The chosen range of rod lengths then corresponds to $4L < L_{P} < 80L$ with small rods being considerably stiffer than long ones.  All rods, long and short, are {\em marginally flexible} and  neither represent rigid bodies $(L_{P} \rightarrow \infty)$ nor semi-flexible chains $(d_{0} < L_{P} < L)$. 

\begin{figure}[htb]
\includegraphics[width=\columnwidth]{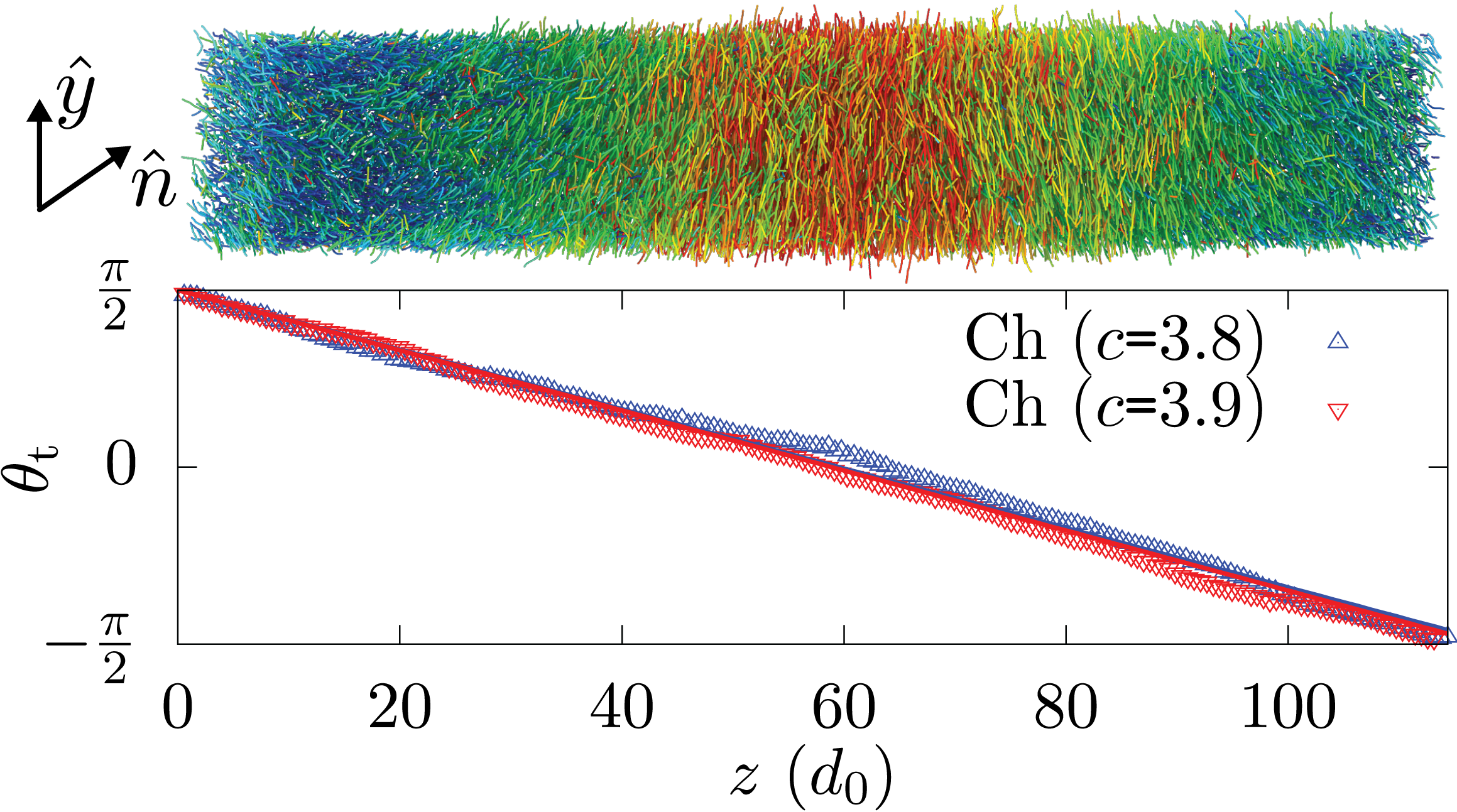}%
\caption{\label{fig:pitch} Twist angle $\theta_{t}  = \cos ^{-1} ({\bf \hat{n} \cdot {\bf \hat{y}}}) $ demonstrating uniform helical twist of the cholesteric phase along the $z$-axis of the simulation box with a normalized pitch length $P/\bar{L}= 16 \pm 5$.}
\end{figure}

\begin{figure}[htb]
\includegraphics[width=\columnwidth]{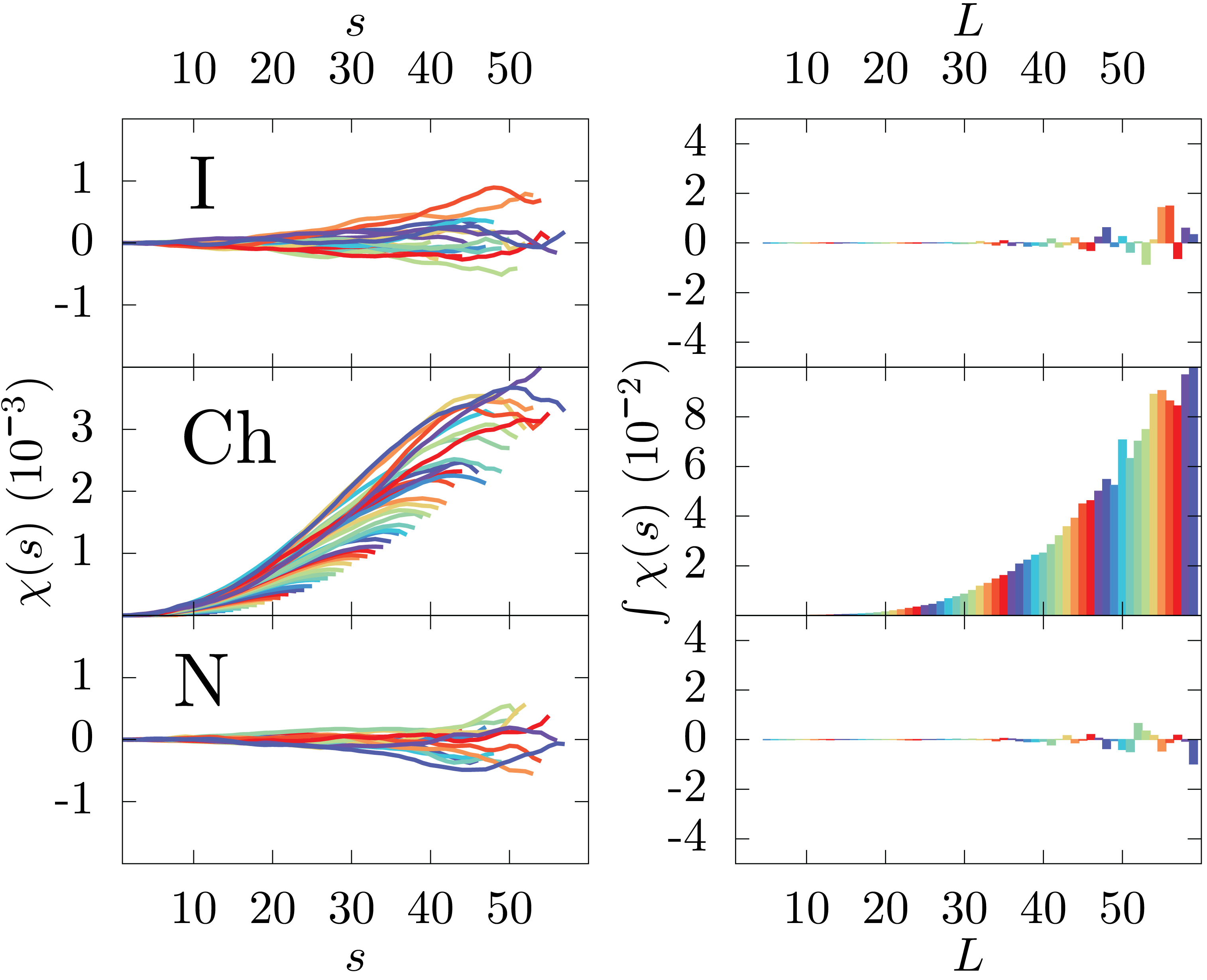}% Here is how to import EPS art
\caption{\label{fig:intra_chiral} Fluctuation-driven shape chirality in near-rigid rods as a function of the subcontour length $s<L$ for a number of rods with length $L$, quantified by a pseudo-scalar chiral order parameter \eq{intrachiral}. The rightmost panels show the integrated chiral strength per rod for the three basic fluid symmetries: isotropic (I), cholesteric (Ch) and nematic (N).}
%The panels on the right the integrated chiral strength for each contour length $L$. Shown are results for the three basic fluid symmetries:  isotropic (I), cholesteric (Ch) and nematic (N).  }
\end{figure}

Simulations are performed using LAMMPS \cite{plimpton1995fast,thompson2022lammps} from an  initial system composed of non-overlapping rods with random positions and orientations at ultralow  concentration $c\approx1.15$ under NVT conditions to obtain the equilibrium pressure $P$. Production runs were performed in the isothermal-isobaric (NPT) ensemble taking an integration time step 0.005~$(\tau)$ in terms of standard MD time  $\tau = (md_{0}^2/(k_\mathrm{B}T))^{1/2}$ with bead mass $m=1.5$  using a Langevin thermostat, with coupling constant $\mathit{\Gamma}=2$ ($\tau^{-1}$) and fixed temperature $T=k_\mathrm{B}/\epsilon=1$ and an isotropic Berendsen barostat with $P_{\mathrm{damp}}=100 \tau$. 
Each system is then compressed at a constant compression rate of  $2\times10^{-7}$ ($\epsilon/d_{0}^{3}/\tau$), until the system resides well into the uniform nematic regime.  A series of NPT runs are then bifurcated at different pressures to sample different state-points across the isotropic (I) and nematic (N) fluid regions. Each run is performed for $10^6\tau$, with one long run for each unique phase identified of $10^7\tau$ to obtain long time statistics.
\begin{figure}[htb]
\includegraphics[width=\columnwidth]{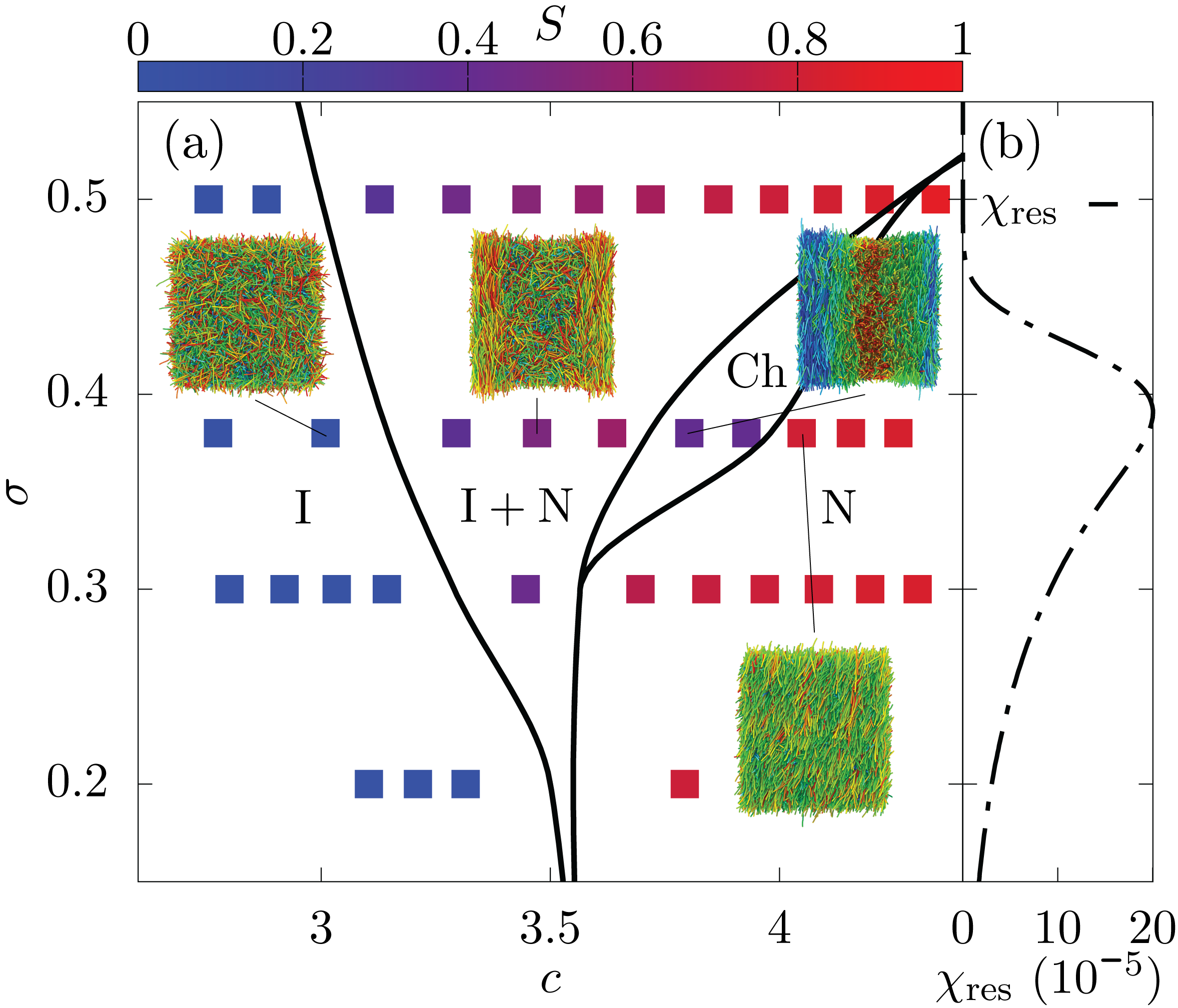}%
\caption{\label{fig:phased} (a) Phase diagram of soft repulsive, near-rigid rods at various length polydispersity $
\sigma$, where nematic order $S$ is indicated by color-coding. The isotropic-nematic phase gap (I+N) widens with polydispersity. A spontaneously twisted cholesteric phase (Ch) emerges at intermediate poydispersity $\sigma  = 0.38$. (b) Residual chirality $\chi_{\rm res}$ peaking at intermediate polydispersity, predicted from the theoretical model [\eq{chires}].}
\end{figure}
The phase diagram shown in \fig{fig:phased} features a characteristic widening of the IN  biphasic region \cite{odijk1985theory} while the concentration at the transition for the least disperse system ($\sigma = 0.2$) roughly corresponds to the prediction  $c \approx 3.29$  from Onsager's theory for shape-uniform rigid rods with $L/d_{0} \rightarrow \infty$ \cite{onsager1949effects}.  The overall degree of  nematic order is measured from the largest eigenvalue $S$ of the second-rank tensor $ Q_{\alpha \beta} =(\sum_{i} 3 \bhu_{i \alpha} \bhu_{i \beta}  - \delta_{\alpha \beta})/2N$ where $\bhu_{i\alpha}$ denotes the $\alpha$-th Cartesian coordinate of the end-to-end unit vector of rod $i$.  The  most striking feature, however, is a sudden drop in global nematic order $S$ at $\sigma = 0.38$  and the emergence of a cholesteric structure with spontaneous helical twisting of the nematic director $\bn(r_{\alpha}) = (\sin qr_{\alpha}, \cos qr_{\alpha}, 0)$ along some arbitrary pitch direction $r_{\alpha}$ of the simulation box. The extent of twist follows from the pitch $P = 2 \pi/q$ which in most cholesteric systems by far exceeds the average rod length.  Indeed, $P/\bar{L} = 16 \pm 5$ for the cholesteric phase depicted in \fig{fig:pitch}  with the  error partly caused by a weak bias exerted by the periodic boundary conditions applied along the pitch axis ($r_{\alpha} = z$) \cite{allen1993computer}.      
Since the rods are composed of strictly achiral, repulsive spherical beads the emergence of director twist  must be driven by a sequence of  chiral shapes each rod adopts as its backbone fluctuates.  Evidence for the existence of transient or long-lived chiral conformations can be obtained  from a correlation function measuring local chirality over each sub-contour of length $s$ along the backbone of rod with total length $L$
\beq
\chi(s)  =\left \langle [{\bf \hat{e}}_{n} \times {\bf \hat{e}}_{m}] \cdot \Delta \hat{\bf r}_{nm} \right \rangle_{| n - m| = s} 
\label{intrachiral}
\eeq
in terms of a pseudo-scalar applied to each bond with position ${\bf p}_{n}= ({\bf r}_{n+1} + {\bf r}_{n})/2$  (with ${\bf r}_{n}$ the position of bead $n$), bond unit vector ${\bf \hat{e}}_{n} = ({\bf r}_{n+1} - {\bf r}_{n}) / ||  {\bf r}_{n+1} - {\bf r}_{n} ||  $ and distance $\Delta \hat{\bf r}_{nm} = ({\bf p}_{m} - {\bf p}_{n}) || ({\bf p}_{m} - {\bf p}_{n}) || $.  The correlation function can be measured for each contour length with the brackets $\langle  . \rangle$ denoting a time average.  If the rods adopt local helical configurations with both left- and right-handed motifs occurring at an equal probability or frequency the above correlation function will average out to zero and the rods are effectively non-chiral. Bearing in mind that there is no unique order parameter defining the chirality of a molecule \cite{lubensky1998chirality} we argue that $\chi$ can be used as a simple but effective measure to quantify chiral shape fluctuations in our rod systems [\fig{fig:intra_chiral}].  

%Indeed, the results in \fig{fig:intra_chiral} confirm the presence of long-lived chiral fluctuations in all phases sampled. The strongest fluctuations are observed in the isotropic fluid where rod crowding is weakest and rods have more space to explore conformational changes than in the denser nematic phase.     

To extract information about the effective backbone shape  we  consider an idealized case where each rod is assumed to coil into a perfect helix parameterized in an arbitrary Cartesian frame as ${\bf p}(s) = \{  R \cos q_{r} s, R \sin q_{r} s, s\}$ ($s \in [0,H]$) with 
$p_{r} = 2 \pi/q_{r}$ the molecular pitch  and $R$ the helix diameter and $H = L/\sqrt{1+ q^{2} R^{2}}$ the  Euclidian  end-to-end distance. Then, from the local tangent vector ${\bf \hat{e}}(s) = d {\bf p}(s)/d s$ we obtain an analytical analog of \eq{intrachiral}
\beq
\chi(s) \sim q_{r}^{5} R^{2} s^{3}
\label{chil}
\eeq
for weakly helical fluctuations $q_{r} R \ll 1$. Tightly wound helical conformations with $q_{r}L \gg 1$ would result in a highly oscillatory $\chi$,  not observed in \fig{fig:intra_chiral},  which suggests that the molecular pitch $p_{r}$ associated with the rod conformations is generally  larger than the contour length.  

While in the isotropic phase $\chi$ remains virtually zero for all sub-contours $s$, indicating that left- and right-handed chiral conformations cancel out over time and rods acquire no net chirality, a uniform positive signal emerges for the cholesteric state with long rods transmitting a systematically stronger chirality than short ones. This hints at spontaneous chiral symmetry breaking (CSB) imparted by the longest rods in the system. Whereas rod shapes are only transiently chiral in the dilute isotropic phase, they become persistently chiral in a more crowded nematic environment. If crowding becomes too strong the effect is much weaker (only noticeable for all but the longest species) and an untwisted nematic (N) is recovered. This is due to a crowding-induced anomalous stiffening of the rods which impairs their ability to effectively transmit chirality through locally chiral conformations along the rod backbone.  

However, the emergence of a stable cholesteric driven by polydispersity cannot be explained on account of a temporal CSB of shape fluctuations alone;  even when rods pick and retain a certain handedness over time, each subsequent rod would still feel no preference for one symmetry over the other and a racemic mixture would be expected for every species $L$, leading to zero net chirality.  The results in \fig{fig:intra_chiral} indicate that this is not the case for the cholesteric state where  $\chi > 0$ systematically for all longer-than-average rods. This suggests the presence of {\em chiral synchronization} of particle conformations \cite{tschierske2016mirror} giving rise to  chiral interparticle forces and spontaneous twist at the macroscale [\fig{fig:pitch}]. 

Intriguingly, the cholesteric structure is only found at intermediate length dispersity $\sigma = 0.38$ and does not show up for the other values explored (\fig{fig:phased}). This suggests that rod length dispersity  plays a subtle but crucial role in the CSB observed here. Herein, this new finding is rationalized using a simple model based on the perfect helix introduced previously. Firstly note that not all helical conformations are energetically equivalent but are distributed according to the bending energy of a worm-like chain, leading to a probability density $f$ 
\beq
f(q_{r}R) \propto \exp \left (  -\frac{L_{P}}{2} \int_{0}^{H} ds (\partial_{s} {\bf p})^{2} \right )
\eeq
For stiff rods $L_{P} \gg L$ and weak curliness $q_{r}R  \ll 1$ the result is the distribution
$f(q_{r}R)  \propto \mathcal{N} e^{- \frac{1}{2}L_{P} L q_{r}^{2}(q_{r}R)^{2}}$ that is Gaussian in $q_{r}R$ with $\mathcal{N}$ a normalization factor.  Next, a chiral propensity $W(L)$ is defined from averaging over all helical conformations of a single handedness 
\begin{align}
W(L) & =  \int_{0}^{\infty} d(q_{r}R) f(q_{r}R) \chi(L) \sim  -\frac{L^{2}q_{r}}{ L_{P}}  
\end{align}
which tells us that the helical fluctuations of a long rod have a stronger propensity to impart chirality than those of short ones.  Let us assume $N_{L}$ rods of length $L$  each adopting a RH (LH) symmetry with probability $p$ ($1-p$). Once a rod has picked a certain handedness it will retain it over time reflecting temporal CSB as ascertained from our simulations. The average chiral propensity of a rod with length $L$ is given by
\beq
\langle W(L) \rangle = \sum_{k=1}^{N_{L}} \left \{ P_{k}(N_{L}) k - P_{N_{L}-k}(N_{L})(N_{L}-k) \right \} W(L)
\eeq
with $P_{k}(N)=\binom{N}{k} p^{k}(1-p)^{N-k}$ is the binomial distribution. Let us focus on the most conservative case of {\em no chiral synchronization} reflected by $p = 1/2$, i.e., each rod adopts a LH or RH shape with equal probability. Then, the propensity simplifies to 
$\langle W(L) \rangle \sim -q_{r}L^{2}N_{L}2^{-N_{L}}/L_{P}$ with the brackets denoting an ensemble average over $N_{L}$.
Clearly, the expected chiral propensity rapidly drops to zero for $N_{L} \gg 1$. As expected, for a monodisperse system this simply gives a racemic mixture with  {\em no net chirality} in the thermodynamic limit $N \rightarrow \infty$. However, for a length-disperse system defined by some fat-tailed distribution $p(L)$ (e.g. log-normal, see \fig{fig:dist}) the system is too deficient in very long rods to reach the racemic limit. The last step is then to average  over all species with normalized length $\ell  = L/\bar{L}$ to obtain  a measure for the  {\em residual chirality} $\chi_{\rm res}$ of the whole system
\beq
\chi_{\rm res} =  \overline{ \langle W(L) \rangle} \sim -\frac{q_{r} \bar{L}^{2}N}{L_{P}} \int_{0}^{\ell_{m}} d \ell p(\ell)^{2} \ell^{2} 2^{-N p(\ell)} 
\label{chires}
\eeq
where the overbar denotes an average over all species. 
The residual chirality depends on the system size $N$ and the shape of $p(\ell)$ and most notably the tail cut-off $\ell_{m} \gg 1$. For bounded distributions with finite $\ell_{m}$ net chirality is zero, since $\lim_{N \rightarrow \infty} \chi_{\rm res} =0$. The explanation is quite simple; there are too many rods with maximum length  randomly adopting  long-lived RH or LH shapes so that a racemic mixture is obtained. An interesting case, however, arises when both $N$ and  $\ell_{m}$ simultaneously tend to infinity. The two variables can be linked by considering  rods in a cubic (simulation) box of size $L_{b}$ and volume $V = L_{b}^{3}$. Assuming the longest species to be half the box size $\ell_{m}  = L_{b}/2\bar{L}$. Then, the total number of rods scales as $N = (32/\pi) c (\bar{L}/d_{0}) \ell_{m}^{3}$ with average aspect ratio $\bar{L}/d_{0} = 15$ and $c \approx 4$ the relevant dimensionless concentration in the phase diagram [\fig{fig:phased}]. Plugging in the log-normal distribution, $\chi_{\rm res}$ is strongly non-monotonic and peaks at intermediate $\sigma$ while being virtually zero for near monodisperse as well as for widely polydisperse systems. This is in complete agreement with the phase diagram where the cholesteric phase only appears at intermediate $\sigma = 0.38$, see the rightmost panel of Fig  \ref{fig:phased}.

In conclusion, we  demonstrate that two key attributes of many biopolymers namely backbone flexibility and length polydispersity  can conspire to generate liquid crystals  with spontaneously broken chiral symmetry even when the particles are not natively chiral and do not otherwise feature any broken particle symmetry.  While achiral shapes such as bent-core molecules are known to form so-called twist-bend liquid crystals, we report the formation of  stable cholesteric structures formed from strictly achiral  near-rigid rods provided  the particles exhibit a sufficiently wide spread in contour lengths. The subtle role of size polydispersity in driving spontaneous mirror symmetry breaking at the macroscale has not been noted before and is illustrated by a simple theoretical model that predicts residual chirality imparted by longer-than-average rods to emerge only at intermediate size polydispersity.  Our model is generalizable to a wide variety of lyotropic biopolymer LCs, composed of non-rigid moieties, where shape  dispersity  can lead to enhanced chirality transmission between natively chiral biological nematogens. 
Such dynamic modes of polydispersity-driven CSB, in moieties with transient conformational chirality, could be harnessed in the future to obtain homochiral agents, pharmaceutical ingredients and materials.

This work was supported by the European Innovation
Council (EIC) through the Pathfinder Open grant “INTEGRATE” (no. 101046333). The authors acknowledge HPC resources from GENCI-IDRIS (Grant 2024-[A0170913823]).

%\begin{figure}
%	\includegraphics[width = \columnwidth]{prop_ss}
%	\caption{Residual chirality $\chi_{\rm res}$ in units $q_{r}\bar{L}^{2}/L_{P}$ for a length-disperse system of near-rigid rods as a function of the  polydispersity $\sigma$ for a log-normal and Schulz distributions.  \red{If we manage to get decent data for the typical correlation time of $\chi$ going from I-Ch-N we could add them here. I suggest only showing log-normal curve, not Schulz. Data file  "chires.dat" and Mathematica source code "tailgate.nb" added to Overleaf.  } }
%	\label{happy2}
%\end{figure}

\bibliographystyle{apsrev4-1}
\bibliography{fall_et_al}

\end{document}